\documentclass[aps,a4paper,nofootinbib,superscriptaddress,
tightenlines,floats,amsfonts,amssymb]{revtex4}

\usepackage[dvips]{graphicx}
\usepackage[dvips]{epsfig}

\def\sin{{\rm sin}}
\def\cos{{\rm cos}}
\def\tan{{\rm tan}}

\def\be{\begin{equation}}
\def\ee{\end{equation}}
\def\bea{\begin{eqnarray}} 
\def\eea{\end{eqnarray}}

\def\simleq{\; \raise0.3ex\hbox{$<$\kern-0.75em
      \raise-1.1ex\hbox{$\sim$}}\; }
\def\simbeq{\; \raise0.3ex\hbox{$>$\kern-0.75em
      \raise-1.1ex\hbox{$\sim$}}\; }
\begin{document}

\title{Comment on ``Gravitating magnetic monopole in the
global monopole spacetime''}
\author{A.~Ach\'ucarro} \affiliation{Lorentz Institute for Theoretical
Physics, University of Leiden, The Netherlands}
\affiliation{Department of Theoretical Physics, UPV-EHU, Bilbao,
Spain} \author{J.~Urrestilla} \affiliation{Department of Theoretical
Physics, UPV-EHU, Bilbao, Spain}

\date{\today}

\begin{abstract}
We point out a problem with the stability of composite
(global--magnetic) monopoles recently proposed by J.~Spinelly,
U.~de~Freitas and E.R.~Bezerra~de~Mello [Phys.~Rev.~{\bf D66}, 024018
(2002)].
\end{abstract}

\maketitle

Spinelly et.~al. \cite{SBF} have written an interesting paper on {\it
composite} monopoles, i.e., a global monopole and a magnetic monopole
bound together by their gravitational interaction. This paper was
followed by two works considering more general cases \cite{BH,BBH}.
Such {\it composite} configurations are of interest for several
reasons

One is in connection with the (magnetic) monopole problem.  Global
monopoles have long range interactions, and their annihilation rate is
very efficient. Composite monopoles could share this property, if they
remained bound as the global monopoles move to annihilate each other.
One could even imagine a scenario in which the global and magnetic
monopoles are created independently, with the magnetic monopoles
coming from an earlier phase transition. A second transition leads to
global monopoles which then capture and drag the magnetic monopoles
with them as they go towards a neighbouring defect.  Therefore
magnetic monopole annihilation could be enhanced with respect to the
case of purely magnetic monopoles. If there is no correlation between
the signs of the global and magnetic charges, the effect is quite mild,
and gives rise to a remnant population of doubly charged, purely
magnetic monopoles.  On the other hand, if the global and magnetic
charges {\it are} correlated the effect could be dramatic, strong enough to
solve the monopole problem.

Another reason is the spacetime structure around such defects.  Pure
global monopoles cause a deficit solid angle in the surrounding 
spacetime\footnote{The area of a sphere of radius $r$ surrounding the
monopole is $4\pi (1 - \Delta) r^2$} $\Delta$, plus a small effect due to the
core \cite{BV}. The mass that can be ascribed to the core turns out to
be negative \cite{HL}, leading to repulsive interactions. The presence of
a magnetic monopole component can in principle change the sign of the
core mass to make it positive \cite{SBF}. Finally, the critical
phenomena associated with gravitational collapse of the monopole to
form a black hole show an interesting intermediate behaviour between
that of the global and magnetic cases \cite{BH}.

Our comment consists basically of two points. First, some of the
results of \cite{SBF} have a very simple physical interpretation that
was somewhat obscured by their choice of units and that we want to
clarify. Second, and more important, we want to emphasise that the
composite monopole that is bound only gravitationally is almost
certainly unstable to breaking up into its global and magnetic
monopole constituents; therefore, if it is to be cosmologically
relevant some stronger type of interaction must be introduced between
the two sets of fields. Non-minimal coupling has been considered in
\cite{BBH}, but only in spherically symmetric situations, and this
problem was not addressed. More work will be needed in order to
establish the existence of {\it stable} composite monopoles.

We begin by clarifying the interpretation behind the apparent
dependence of the global monopole field profile with the parameter
$\beta$, shown in figures 3a) and 6c) of ref. \cite{SBF}, which is
just an artifact of their choice of units.

Considering first the case where there is only a gravitating global
monopole, we see that there are two mass scales in the problem. One is
set by the v.e.v of the scalar field, $\eta$, the other by the massive
scalar excitations, which have mass $\sqrt \lambda\eta$ ($\lambda$ is
the quartic coupling). We can use
the second one to define a natural length scale, and the first one
will define an overall scale for the energy. In other words, in the
absence of a magnetic monopole, the rescaling 
\be \chi^a\to\eta
\chi^a\,,\qquad x_\mu\to {\hat x}_\mu =
\frac{x_\mu}{\eta\sqrt{\lambda}}\,,
\label{res} 
\ee 
in Lagrangian 
(7),(10) in reference~\cite{SBF} makes the parameter $\lambda$ disappear from the equations of
motion, while the parameter $\eta$ appears only in the adimensional
combination $G\eta^2$, which measures the gravitational strength of
the global monopole (and, in particular, the deficit solid angle
$\Delta = 8\pi G\eta^2$ in the spherically symmetric
configuration). The spherical monopole profile $f(\hat r)$ 
is known to be fairly independent of $\Delta$ \cite{HL},
and we expect this feature to persist in the presence of the magnetic
monopole; this independence is observed in ref. \cite{SBF}.

\begin{figure*}[!htb]
\begin{center}
\includegraphics[angle=-90, width=7cm]{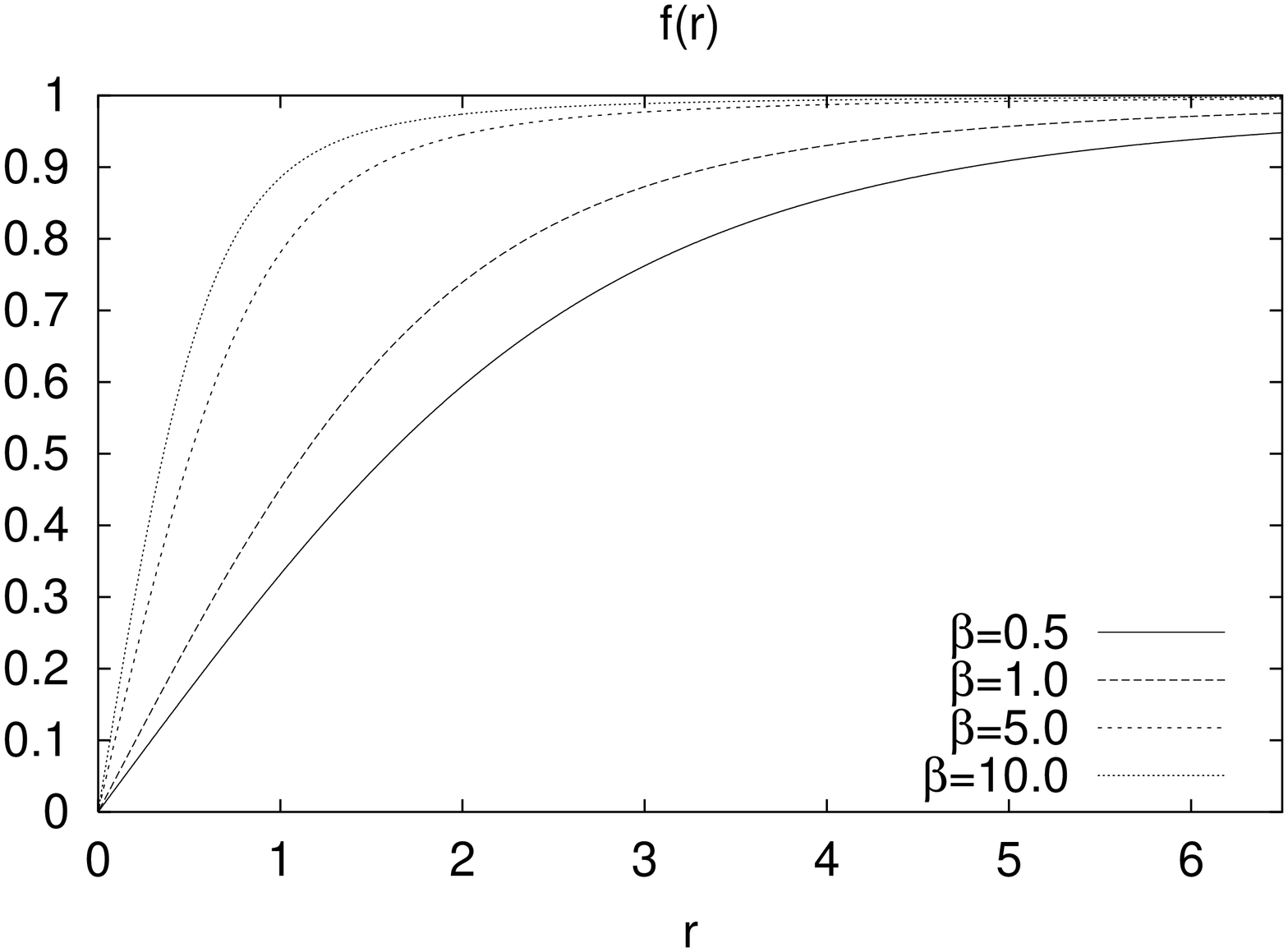}
\hspace{1cm}
\includegraphics[angle=-90, width=7cm]{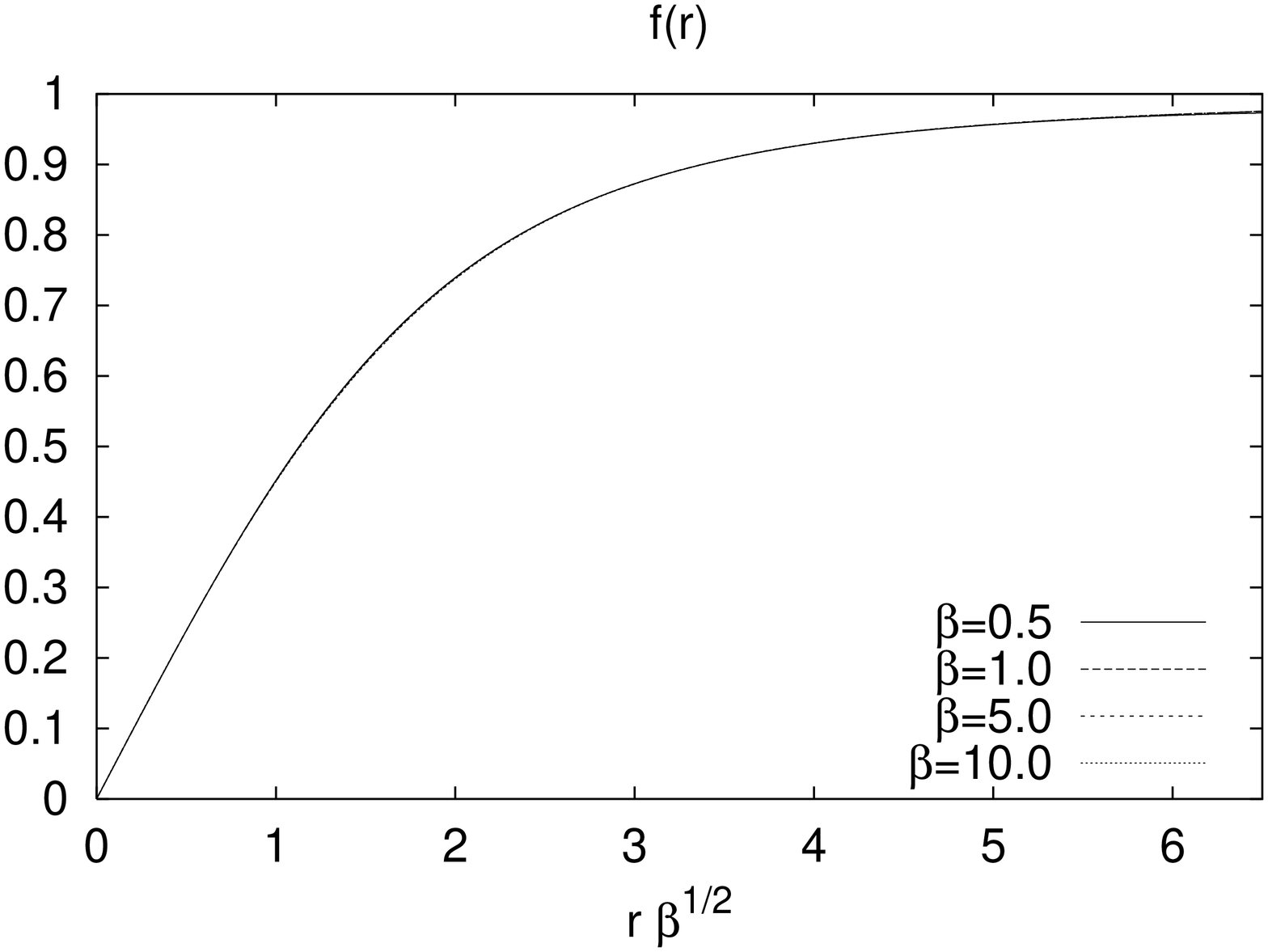}
\caption[lambda]{\label{lambda} The function $f(r)$ for
different values of $\beta$, with $\Delta=0.1$ fixed. The figure in the
right column shows that choosing the inverse of the scalar mass as the
unit of length, the global monopole profile $f(r)$ is insensitive to
the value of $\beta$.}
\end{center}
\end{figure*}

In general, the magnetic monopole would introduce three
new mass scales: the v.e.v. of its scalar field, the mass of the
scalar excitations and the mass of the vector boson, $e\eta$ ($e$ is the 
gauge coupling constant). In
ref. \cite{SBF} the first two were chosen to coincide with those of
the global monopole, so the only new adimensional
parameter is the ratio of the scalar mass to the vector mass. The
parameter $\beta = \lambda / e^2$ measures this ratio (squared).
There are no more free parameters in the problem.

We do not expect the presence of the magnetic monopole to alter the
global monopole profile $f({\hat r})$ substantially, since there is no
direct coupling between them. In particular $f$ should be quite
insensitive to the value of $\beta$, if $\Delta$ is kept fixed.  The
only reason why figs. 3a) and 6c) in ref. \cite{SBF} 
appear to give such a dependence is
that in \cite{SBF} the radial coordinate was rescaled using the
{\it vector} boson mass instead of the scalar mass (a desirable choice
in other respects). The difference is just a factor of $\sqrt \beta$
in the abscissa and once this is taken into account the curves agree
perfectly, as shown in figure \ref{lambda}.

We now turn to the composite monopole's stability.
To understand the nature of the problem,
consider the simpler case of the interaction between a magnetic
monopole and a conical singularity (such as would be produced, for
example, by 
an infinitely thin, straight, idealised cosmic string). 
In
the plane perpendicular to the string, the metric is flat and has no
effect on test masses other than lensing. But this is not true of
gravitating masses and charges, whose long--range electric or magnetic
fields can detect the presence of the singularity.  The net effect
turns out to be that masses are attracted to the conical singularity
and charges (electric and magnetic) are repelled \cite{Smith}. In the case of a
magnetic charge, the convergence of magnetic field lines behind the
singularity is responsible for the repulsion.

Now consider the three-dimensional version, a magnetically charged 
point particle in a
spherically symmetric spacetime with a deficit {\it solid} angle
concentrated in a point-like singularity. Without loss of generality
the charge can be assumed to move in the equatorial plane
($\theta = \pi/2$).

This is a reasonable approximation to the interaction between a global
monopole and a magnetic monopole that are well separated.
But then the arguments of
\cite{Smith} apply, and we expect the monopoles to repel.
Moreover, in this case, even the gravitational interaction is
repulsive, as the ``mass'' of the core of the global monopole is
negative. An explicit calculation of the interaction between a pointlike {\it
electrically} charged particle and a global monopole (including the effect of the core) 
confirms that it is
repulsive \cite{BF}; we expect this result to hold when considering 
a magnetic
charge and a global monopole.

We conclude that the
cosmological capture of a magnetic monopole by a global monopole seems
quite improbable (we have neglected the effect of the finite--size
cores in this argument, but since the monopoles only interact
gravitationally we do not expect our conclusions to change qualitatively).  While
this is disappointing, it is not necessarily a problem for the composite
monopole since in that case the cores are superposed, the magnetic
field lines are evenly distributed with spherical symmetry and thus
there is no net repulsion between the two cores.

The problem arises when one considers the stability of the composite
monopole to {\it angular} perturbations. This is an important
consideration because, in flat space, global monopoles have a zero mode which allows
the redistribution of gradient energy density in an axisymmetric way,
without any cost in energy \cite{G}. Following the notation in
\cite{SBF}, where the spherical global monopole is given by:
\be
\chi^a=f(r)\frac{x^a}{r}\,,
\ee
purely angular deformations with axial symmetry can be described using the following ansatz
\cite{G}
\bea
\chi^1& =& f(r)\, \sin\,\bar{\theta}(r,\theta)\, \cos\,\varphi\,; \nonumber\\ 
\chi^2& =& f(r)\, \sin\,\bar{\theta}(r,\theta)\, \sin\,\varphi\,; \nonumber\\ 
\chi^3& =& f(r) \, \cos\,\bar{\theta}(r,\theta)\,.
\label{ansatzgoldhaber}
\eea
Note that $\bar\theta=\theta$ corresponds to the unperturbed monopole.

As shown by Goldhaber \cite{G}, in flat space, configurations of the form
\bea
\tan\left({{\bar \theta} \over 2}\right)=\tan\left({\theta \over
 2}\right) e^\xi\,, \qquad \xi={\rm const} \,,
\label{conf}
\eea
where $\xi$ is an arbitrary constant, are degenerate in energy with the spherical monopole.
For any given $r$, $\xi\gg 1$ corresponds to concentrating the gradient energy in an arbitrarily small region
 around the north-pole  {\it with no cost in energy}.

The extra tension created at the north pole drags the core of
the global monopole upwards, restoring spherical symmetry
\cite{BR,P,AU,WT}. But consider its effect on the magnetic monopole: a
larger gradient energy density in the global monopole causes a
redistribution of the deficit angle ``density'', with the larger
contribution now coming from the north pole. But then magnetic field
lines are pushed together there, causing the magnetic core to move
{\it downwards}, i.e. in the opposite direction to the global monopole
core. Once the cores are separated, they are expected to repel, so an
attractive interaction between the cores seems {\it necessary} for the
composite object to survive.

To summarise, the stability of composite monopoles can by no means be
taken for granted and is expected to depend quite sensitively on the
details of the interaction between the two constituents.  Any attempt
to consider composite monopoles in an astrophysical or cosmological
context must address this problem.

\begin{acknowledgments}
We are grateful to E.R. Bezerra de Mello for
pointing out ref. \cite{BF}. J.U. thanks the University of Leiden, where part of this
work was done, for hospitality. This work is partially supported by 
the ESF COSLAB programme, and by AEN99-0315, FPA 2002-02037
 and 9/UPV00172.310-14497/2002 grants.
\end{acknowledgments}

\end{document}